\documentclass[12pt,superscriptaddress]{article}

\pdfoutput=1
\usepackage{jheppub}
\usepackage{epsfig}
\usepackage{amsmath,amssymb}
\usepackage{amsfonts}
\usepackage{mathrsfs}
\usepackage{units}

\newcommand{\beq}{\begin{equation}}
\newcommand{\eeq}[1]{\label{#1}\end{equation}}
\def\beqa{\begin{eqnarray}}
\def\eeqa#1{\label{#1}\end{eqnarray}}
\newcommand{\eeqn}{\end{equation}}
\newcommand{\CR}{\notag \\}
\newcommand{\leqn}[1]{\eqref{#1}}

\def\to{\rightarrow}
\def\tr{{\rm tr}}

\newcommand{\sptwo}{B{}}           % Spin two tensor
\newcommand{\warp}[1]{\ensuremath{e^{#1 k y}}}  % The warping factor that shows up from the metric
\newcommand{\ud}{d}              % d for derivatives and infinitesmals
\newcommand{\kkn}[1]{#1^{(n)}}            % For KK modes, functions etc
\newcommand{\kkm}[1]{#1^{(m)}}            % For KK modes, functions etc
\def\LIR{\Lambda_{\rm IR}}

\def\stacksymbols #1#2#3#4{\def\theguybelow{#2}
    \def\vp{\lower#3pt}
    \def\sp{\baselineskip0pt\lineskip#4pt}
    \mathrel{\mathpalette\intermediary#1}}

\def\intermediary#1#2{\vp\vbox{\sp
     \everycr={}\tabskip0pt
     \halign{$\mathsurround0pt#1\hfil##\hfil$\crcr#2\crcr
              \theguybelow\crcr}}}

\begin{document}

\begin{center}
%{\Huge \bf DRAFT - PRELIMINARY!!!}
\end{center}
\vskip1.5cm
\title{Four boosted tops from a Regge gluon}
\author[a]{Maxim Perelstein,}  
\emailAdd{mp325@cornell.edu}
\affiliation[a]{Laboratory of Elementary Particle Physics, Physical Sciences Building, Cornell University, \\ Ithaca, NY 14853, USA}
\author[b]{Andrew Spray} 
\emailAdd{aps37@triumf.ca}
\affiliation[b]{TRIUMF, \\ 4004 Wesbrook Mall, Vancouver, BC V6T 2A3, Canada}

\vskip.5cm
\date{\today}

\abstract{The hierarchy problem can by addressed by extending the four-dimensional space-time to include an extra compact spatial dimension with non-trivial ``warped'' metric, as first suggested by Randall and Sundrum.  If the Randall-Sundrum framework is realized in string theory, and if the Standard Model particles propagate in the extra dimension, Regge excitations of the Standard Model states should appear around the TeV scale. In a previous publication, we proposed a field-theoretic framework to model the tensor (spin-2) Regge partner of the gluon. Here, we use this framework to study the collider phenomenology of this particle. We find that Regge gluon decays involving Kaluza-Klein (KK) partners of Standard Model fields are very important. In particular, the decay to two KK gluons (with one possibly off-shell) dominates in most of the parameter space.  This decay produces a very distinctive experimental signature: four highly boosted top quarks.  We present a preliminary study of the detection prospects for this signal at the Large Hadron Collider (LHC).  We find that Regge gluons masses up to about 2~TeV can be probed with 10~fb$^{-1}$ of data at 7~TeV center-of-mass energy. With design luminosity at 14~TeV, the LHC should be sensitive to Regge gluon masses up to at least 3.5~TeV.}


\arxivnumber{1106.2171}

\maketitle

\newpage

\section{Introduction}

One of the most promising extensions of the Standard Model (SM) at the TeV scale postulates the existence of an extra compact spatial dimension with a ``warped" (non-factorizable) metric. Randall and Sundrum (RS)~\cite{RS} pioneered this class of models, and demonstrated that warping can naturally produce a radiatively stable hierarchy between the weak and Planck scales. While in the original model the SM matter fields, including the Higgs, were assumed to be confined to a four-dimensional boundary of the five-dimensional ``bulk" space, it was soon realized that models with matter fields free to propagate in the bulk are more realistic, being less constrained by precision electroweak fits. They also offer a number of intriguing features, such as the possibility of gauge coupling unification~\cite{RSgut}, and a novel explanation of the fermion mass hierarchies and the ``flavor puzzle" (the stronger-than-expected experimental bounds on flavor-changing processes at low energies)~\cite{APS}. In addition to models where the full SM is embedded in the bulk, an interesting variation is the ``Higgsless" framework~\cite{HL,CGHST,HLdeloc}, in which the electroweak symmetry breaking is achieved by imposing non-trivial boundary conditions on the (five-dimensional) gauge fields, eliminating the need for the Higgs field. 
For a review of model-building and phenomenological studies in the RS framework, see ref.~\cite{RS_review}

If the RS framework is realized in string theory, the masses of the Regge excitations of the SM fields are expected to be at or near the TeV scale, due to the same effect of the warped metric that produces the weak-Planck hierarchy~\cite{Regge_top,us1,RW}. This opens up the possibility of experimental detection of such Reggeons, perhaps at the CERN Large Hadron Collider (LHC). To facilitate experimental searches for the Reggeons, it is highly desirable to obtain theoretical predictions of their properties, such as masses, spins, decay channels and branching ratios. Ultimately, incorporating Reggeon production and decay in a Monte Carlo event generator is required to conduct a systematic search for these particles in the data. However, detailed first-principles theoretical description of the Reggeons is currently impossible: it is not known how to quantize string theory on the RS background, meaning that the Reggeon scattering amplitudes (or even their mass spectrum) cannot be derived. In ref.~\cite{us1}, we proposed a toy model which, while not rigorously derived from string theory, should capture the phenomenologically important features of the low-lying Reggeons.\footnote{A similar approach to modeling the Regge excitation of the top quark in the RS model has been proposed in~\cite{Regge_top}.} We focused on the first tensor (spin-2) Regge excitation of the gluon, $g^*$, which is a color octet and can be produced at the LHC as an $s$-channel resonance in either $q\bar{q}$ or $gg$ collisions. The model was constructed as follows:

\begin{itemize}

\item Start with the familiar Veneziano amplitudes, which describe the effect of Reggeon exchanges in scattering among SM states in 4 dimensions, or in models with flat (toroidal) extra dimensions and SM states propagating on a 4-dimensional brane~\cite{CPP,flat_strings}.

\item Obtain the Feynman rules by factorizing the Veneziano amplitudes on their Regge poles (this was first done, for ``stringy quantum electrodynamics", in ref.~\cite{CPP}.)

\item Write down the Lagrangian which produces these Feynman rules, in which the spin-2 Reggeon is described as a massive tensor field.

\item Introduce metric factors into this Lagrangian to implement general covariance in a minimal way, and extend it to 5 dimensions.

\item Perform Kaluza-Klein (KK) decomposition of the resulting 5D field theory, and compute Reggeon masses and scattering amplitudes.

\end{itemize}

By construction, in the flat-space limit (when the RS curvature is taken to zero), the results of this approach reduce to the appropriate poles of the Veneziano amplitudes. Of course, it is possible that the 5D theory may contain operators which disappear in the flat-space limit, or do not contribute to the $s$-channel poles of the scattering amplitudes in this limit. Those operators would not be captured by our approach, which is therefore not, strictly speaking, a controlled approximation to any string theory calculation. It does, however, give a well-defined field-theoretic setup which incorporates the basic features 
expected of Reggeons in the RS framework, and can thus be used to provide guidance to experimental searches for these particles.

The toy model has already been used as a basis for a broad-brush study of Reggeon collider phenomenology in ref.~\cite{us1} (see also~\cite{AGHT}). We computed the tensor Regge gluon production cross section at the LHC, and found that significant production rates are possible: for example, a 2 TeV $g^*$ has a production cross section of about a pb at $\sqrt{s}=14$ TeV. We also estimated the branching ratios of the $g^*$ decays, and concluded that it decays predominantly to $t\bar{t}$. A search for (boosted) top pairs was therefore suggested as the best way to look for the Regge gluon. This part of the analysis, however, was rather over-simplified: only direct decays to SM states were included. In fact, as we will argue in this paper, Reggeon decays involving {\it Kaluza-Klein excitations} of the SM states are equally, or even more, important. This paper will analyze these decays in detail. Among these, the Reggeon decay to two (first-level) KK excitations of the gluon, $g^*\to g^1g^1$, is the most important, with branching ratios as high as 70\%
 in a large part of the parameter space. Even when this decay is kinematically forbidden, a decay  $g^*\to g^1t\bar{t}$, proceeding via an off-shell $g^1$, often has a significant branching ratio. In either case, since $g^1$ predominantly decays to $t\bar{t}$, the end result is a final state with {\it four} boosted top quarks, a highly distinctive signature which can be searched for at the LHC.\footnote{A similar signal in RS models of four \emph{unboosted} tops was considered in Ref.~\cite{Jung:2010ms}.} We will estimate the reach of this search at the current 7 TeV run of the LHC, as well as the future 14 TeV run.

The rest of the paper is organized as follows. Section~\ref{sec:model} briefly reviews the RS model with matter in the bulk, as well as the toy model for Reggeons proposed in~\cite{us1}. Section~\ref{sec:decays} discusses the Regge gluon decays to KK excitations of the SM particles, which were not included in~\cite{us1}. Section~\ref{sec:LHC} contains the estimates of the LHC sensitivity to Reggeon production via a search for four boosted top quarks. We conclude, and discuss some directions for future work, in Section~\ref{sec:conc}.

\section{Reggeon toy model: a brief review}
\label{sec:model}

The toy model of ref.~\cite{us1} is a field theory propagating on the slice of AdS space. We use the metric
\beq
ds^2 = e^{-2ky}\,\eta_{\mu\nu}dx^\mu dx^\nu\,-\,dy^2\,,
\eeq{RS}
with $y\in [0, \pi R]$. The ``ultraviolet" (UV) brane is at $y=0$, while the ``infrared" (IR) brane is at $y=\pi R$. The scale $k$ is of order (though somewhat below) the usual 4-dimensional Planck scale $M_{\rm Pl}$. Solving the gauge hierarchy problem requires that the combination
\beq
\LIR = k \, e^{-k\pi R}
\eeq{LIR_def}
be of order TeV. While the geometry of the model is defined by two variables, $k$ and $R$, most quantities of interest for TeV phenomenology are determined almost exclusively by $\LIR$. In the numerical work of this paper, we will fix $k/\LIR=10^{15}$ for concreteness, but the results depend only very weakly on this choice. 

Following refs.~\cite{ADMS,HL}, we assume that all SM fields (except, possibly, the Higgs) are zero-modes of 5-dimensional (5D) fields that propagate on the entire space. In the gauge sector, we focus on the $SU(3)$ field $A^a_M(x,y)$, choose the gauge $A_5=0$, and impose Neumann boundary conditions on the branes, $\partial_y A_\mu=0|_{y=0,\pi R}$. The KK decomposition of the gauge field is
\beq
A_\mu(x,y) = \frac{1}{\sqrt{\pi R}}\,\sum_{n=0}^{\infty} A_\mu^{(n)}(x)\,\chi^{(n)}(y)\,,
\eeq{KKgauge}
where the ``wavefunctions" $\chi^{(n)}$ are obtained by solving the appropriate equations of motion~\cite{DHR}, and turn out to be simple Bessel functions for $n\geq 1$. (The zero-mode is flat.) They are normalized according to
\beq
\frac{1}{\pi R} \int_0^{\pi R} dy\, \chi^{(n)}(y)\, \chi^{(m)}(y)\,=\, \delta_{mn}. 
\eeq{gaugenormal}
The spectrum of the KK modes is also obtained from the equations of motion. The KK masses are given by $m(g^l) = C_l \LIR$, where $C_l$ are numerical coefficients: for example, $C_1\approx 2.4$. 

Each SM Weyl fermion $q_{i\sigma}$, where $i \in (u,d,s,c,b,t)$ is the flavor index and $\sigma \in \{L, R\}$ is the chirality, is embedded as a zero-mode of a 5D fermion field $Q_{i\sigma}$. The boundary conditions for the 5D field on the branes are chosen so that the zero-mode has the appropriate chirality~\cite{GN,GP}.  Note that the subscript $\sigma$ on the 5D field is simply a label indicating the chirality of the zero-mode; the 5D field itself of course has no chirality. The KK decomposition for the 5D fermions is
\beqa
Q_{iL}(x,y) &=& \frac{1}{\sqrt{\pi R}} \varphi^{(0)}_{iLL}(y) P_L q_i(x) + \frac{1}{\sqrt{\pi R}}\,\sum_{n=1}^{\infty} \left( \varphi^{(n)}_{iLL} (y) P_L + \varphi^{(n)}_{iLR} (y) P_R\right) q^{(n)}_{iL} (x) \,,\CR
Q_{iR}(x,y) &=& \frac{1}{\sqrt{\pi R}} \varphi^{(0)}_{iRR}(y) P_R q_i(x) + \frac{1}{\sqrt{\pi R}}\,\sum_{n=1}^{\infty} \left( \varphi^{(n)}_{iRL} (y) P_L + \varphi^{(n)}_{iRR} (y) P_R\right) q^{(n)}_{iR} (x) \,,\CR
\eeqa{KKfermi}
where $q_i$ is the massless 4D {\it Dirac} fermion in which the chiral zero-modes of $Q_{iL}$ and $Q_{iR}$ are combined, while $q_{iL}^{(n)}$ and  $q_{iR}^{(n)}$ are massive 4D Dirac fermions, KK excitations of the fields $Q_{iL}$ and $Q_{iR}$ respectively. The normalization condition
\beq
\frac{1}{\pi R} \int_0^{\pi R} dy\, e^{-3ky} \varphi^{(n)*}_{i\sigma_1\sigma_2} (y)\, \varphi^{(m)}_{i\sigma_1\sigma_2} (y)\,=\, \delta^{mn}\,,
\eeq{ferminormal}
imposed for all $\sigma_1$, $\sigma_2$ and $i$ (no summation over these indices is implied), ensures that the 4D fermion fields are canonically normalized.  
While the gauge wavefunctions are completely fixed by the geometry ($k$ and $R$), the fermion wavefunctions depend on the 5D mass $M_{i\sigma}$, which enters via the dimensionless ratio $c_{i\sigma} = \pm M_{i\sigma}/k$. (We take the positive (negative) sign for fields with right (left) handed zero modes.) Note that the existence of a massless zero-mode is independent of $c_{i\sigma}$, but the shape of the zero-mode wavefunction (which in turn dictates its interaction strength with KK gauge bosons, Reggeons, etc.) does depend on $c_{i\sigma}$, via
\beq
\varphi^{(0)}_{i\sigma\sigma}(y) = \sqrt{\frac{\pi k R (1-2c_{i\sigma})}{e^{\pi k R (1-2c_{i\sigma})} - 1}}\,e^{(2-c_{i\sigma})ky}\,,
\eeq{fermi_zero}
and so does the spectrum of KK-excited fermions. 

In this paper, we will consider two sets of 5D fermion masses. The first set, which we will call ``Model A", comes from a model with a light Higgs localized at or near the IR brane~\cite{ADMS}: 
\begin{align}
c_{Q^1} &\approx 0.63, & c_{u^1} & \approx 0.675, & c_{d^1} & \approx
0.675\,,\notag \\
c_{Q^2} &\approx 0.575, & c_{u^2} & \approx 0.5, & c_{d^2} & \approx
0.64\,, \label{cs_Higgs} \\
c_{Q^3} &\approx 0.39, & c_{u^3} & \approx -0.19, & c_{d^3} & \approx
0.62\,. \notag
\end{align}
In this scenario, the first two generations of quarks are mostly elementary, and their couplings to the tensor Reggeon are exponentially suppressed (numerically, the suppression factor is of order $10^{-5}$--$10^{-6}$). The couplings to the third generation doublet and the right-handed top quark are unsuppressed.
In this model, precision electroweak constraints typically imply the KK mass scale (the mass of the first KK excitations of the SM gauge bosons) about 2 TeV or higher~\cite{ADMS, Carena:2006bn}.

The second set, which will be referred to as ``Model B", was originally proposed in the context of the Higgsless model~\cite{CGHST,HLdeloc}:
\begin{align}
c_{Q^1} &\approx 0.5, & c_{u^1} & \approx 0.5, & c_{d^1} & \approx
0.5\,, \notag \\
c_{Q^2} &\approx 0.5, & c_{u^2} & \approx 0.5, & c_{d^2} & \approx
0.5\,, \label{cs_HL} \\
c_{Q^3} &\approx 0.39, & c_{u^3} & \approx -0.19, & c_{d^3} & \approx
0.5\,. \notag
\end{align}
In this case, all fermions have a significant composite admixture. While this choice appears somewhat artificial from theoretical point of view, a strong phenomenological advantage is a drastic suppression of the contribution to the $S$-parameter, meaning that models with KK mass scale below 1 TeV can be consistent with precision electroweak fits~\cite{HLdeloc, Foadi:2004ps,MS}.

\begin{figure}[tb]
\begin{center}
\centerline {
\includegraphics[width=4in]{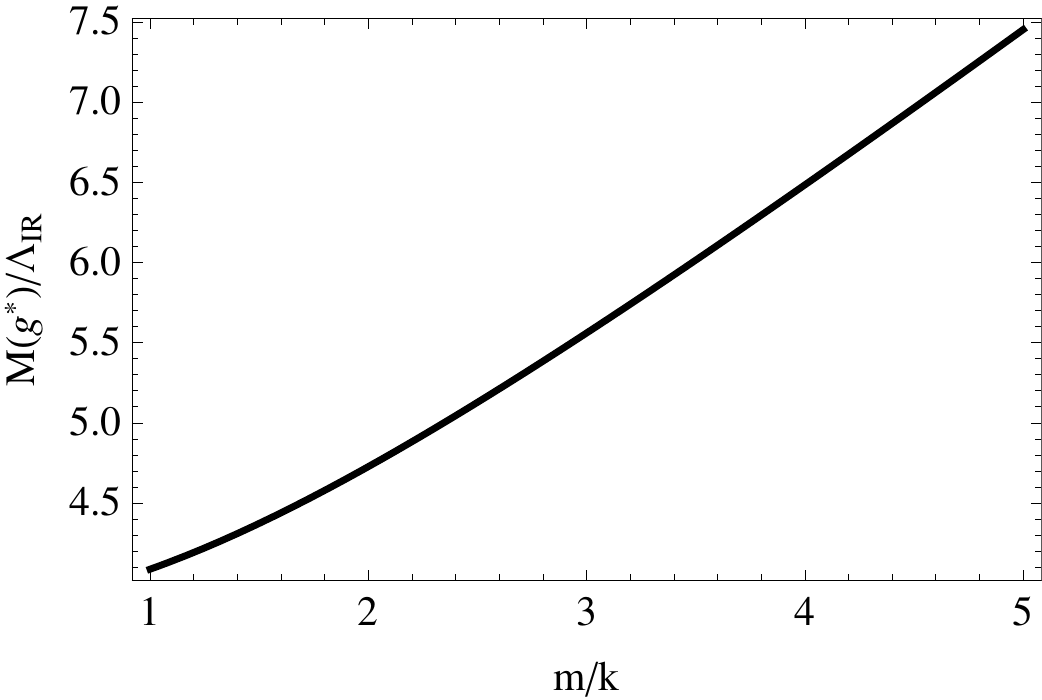}
}
\vspace{-.2in}
\caption{The Reggeon mass in units of $\LIR$.}
\label{fig:Rmass}
\end{center}
\end{figure}

To model the lowest-lying Regge excitation of the gluon in this framework, we introduce a 5D spin-2 field $B_{MN}$, with a 5D mass $m$.\footnote{Note that if $m$ is identified with the string scale, $m\gg k$ would be required to consider the model in terms of a field theory propagating on a smooth 5D geometric background. However, there may be order-one or even larger model-dependent factors entering the relation between $m$ and the string scale. In this paper, we will generally consider $m/k> 1$, but will not assume a large hierarchy between the two scales.}  The KK decomposition of a massive tensor field on the background~\leqn{RS} is non-trivial; this issue was considered in detail in ref.~\cite{us1}. The 5D field decomposes into KK towers of 4D tensor, vector and scalar fields; we focus on the tensor field, since it would provide the most striking signature of the stringy physics involved. The KK decomposition of this field is
\beq
\sptwo_{\mu\nu} (x,y) = \frac{1}{\sqrt{\pi R}} \, \sum_{n=1}^\infty \kkn{\sptwo_{\mu\nu}} (x) \kkn{f} (y)\,,
\eeq{KKregge}
where the wavefunctions have the form
\beq
\kkn{f} (y) = \frac{1}{N} \left\{ J_\nu \left( \frac{\kkn{\mu}}{\LIR} w \right) + c \, J_{-\nu} \left( \frac{\kkn{\mu}}{\LIR} w \right) \right\}\,.
\eeq{KKregge_wavefn}
Here $\mu^{(n)}$ is the mass of the $n$-th KK mode, $w=e^{k(y-\pi R)}$, and $J_\nu$ is the Bessel function of order $\nu\equiv \sqrt{4+(m/k)^2}$. The constants $c$ and $N$ are determined from the boundary conditions and the normalization condition
\beq
\frac{1}{\pi R} \int_0^{\pi R} \ud y \, \warp{2} \kkn{f} \kkm{f} = \delta^{nm} .
\eeq{reggenormal}
For details, see ref.~\cite{us1}. In this paper, we will restrict our attention to the lowest-lying KK mode of the $B_{\mu\nu}$, which we will refer to as simply ``the Reggeon", and denote by $g^*$. The mass of this particle $M(g^*)\equiv M$ is proportional to $\LIR$, with a numerical factor determined by the first zero of the corresponding Bessel function. Typically, $M/\LIR\sim$ a few; see figure~\ref{fig:Rmass}.

The Lagrangian describing interactions between the Reggeon and gauge/fermion fields was constructed in~\cite{us1} according to the procedure outlined in the Introduction. In 5D, the interaction of the Regge gluon with the color gauge field has the form 
\beq
{\cal S}_{ggg^*} \,=\, \int d^5x \sqrt{-G} \,\frac{g_5}{\sqrt{2}M_S^*}\,C^{abc}\,\left(F^{aAC} F^{bB}_C\,-\frac{1}{4}\,F^{aCD}F^b_{CD}G^{AB}\right)\, B^c_{AB}\,,
\eeq{5Dggg}
where $g_5$ is the 5D gauge coupling constant, related to the 4D strong coupling constant via $g_5= \sqrt{\pi R}~g_s$, and $M_S^*$ is the 5D string scale (of order, but somewhat below, the 4D Planck scale $M_{\rm Pl}$). In principle, $M_S^\ast$ and the Reggeon 5D mass $m$ can vary independently, but in this paper we will always take them to be equal.  The color factor is given by
\beq
C^{abc} \,=\, 2 \left( \tr [t^a t^b t^c] + \tr [t^a t^c t^b] \right)\,,
\eeq{Cfactor}
where the $t^a$'s are fundamental $SU(3)$ generators normalized by ${\rm tr}(t^a t^b)=\delta^{ab}/2$.  The Regge gluon interactions with quarks are given by  
\beq
{\cal S}_{q\bar{q}g^*} = \! \int d^5x \sqrt{-G} \,\frac{i g_5}{\sqrt{2}M_S^*}\,G^{LM}\,E^N_a\, %\sum_i 
\sum_{\substack{i \\ \sigma=L,R}}
\left( (\overline{{\cal D}_M Q_{i\sigma}})\Gamma^a \tilde{\sptwo}_{LN} Q_{i\sigma} - 
\overline{Q}_{i\sigma} \Gamma^a  \tilde{\sptwo}_{LN} {\cal D}_M Q_{i\sigma} \right) \,,
\eeq{5Dgqq}
where $\Gamma^n=(\gamma^\nu, i\gamma^5)$, $E^N_n(y) ={\rm ~diag~}(e^{ky},e^{ky},e^{ky},e^{ky},1)$ is the inverse vierbein, and we defined $\tilde{\sptwo}_{LN}\equiv \sptwo^a_{LN} t^a$.  As explained in ref.~\cite{us1}, the covariant derivatives in~\leqn{5Dgqq} can be replaced with ordinary derivatives if one only deals with on-shell Reggeons, as will always be the case in this paper. Plugging the KK decompositions of all fields into eqs.~\leqn{5Dggg},~\leqn{5Dgqq} results in the 4D interaction Lagrangian, which can then be used to derive the Feynman rules (see next section).   

\section{Reggeon decays: including Kaluza-Klein excitations}
\label{sec:decays}

\begin{figure}[tb]
\begin{center}
\centerline {
\includegraphics[width=4in]{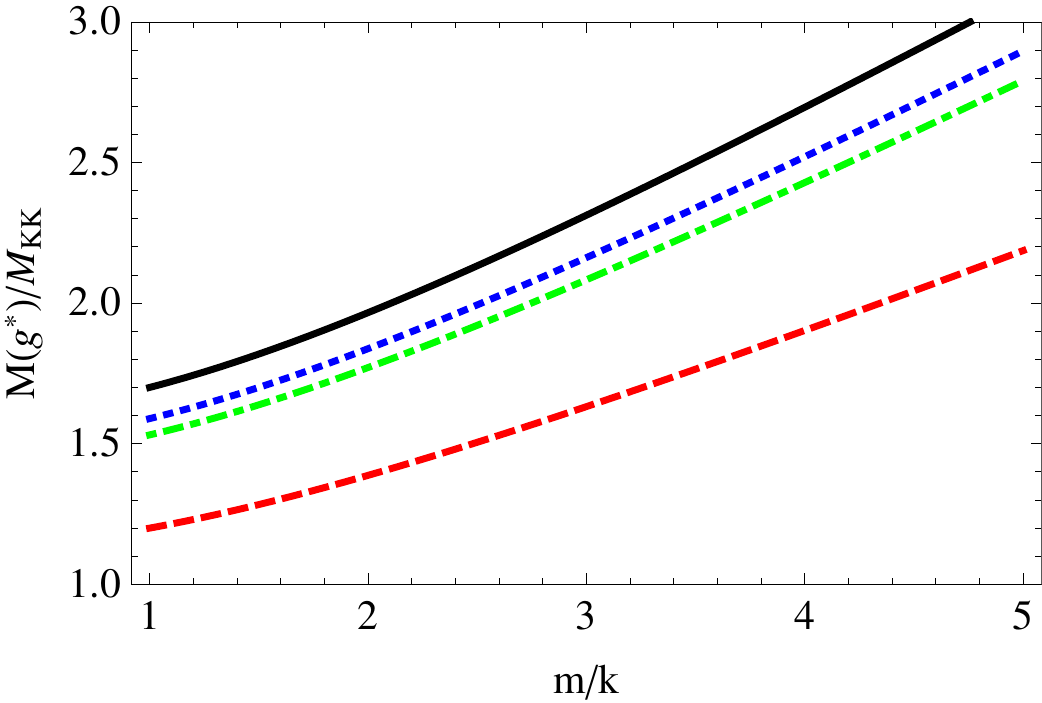}
}
\vspace{-.2in}
\caption{The ratio of the mass of the Reggeon to the $n = 1$ KK states.  The thick, black line corresponds to the KK gluon, as well as to KK quarks with $c = 0.5$.  The other lines are KK quarks for different values of $c$: blue dotted (red dashed, green dot-dashed) correspond to $c = 0.39$ ($-$0.19, 0.67).  The ratios are independent of $\LIR$.}
\label{fig:MRegKK}
\end{center}
\end{figure}

In ref.~\cite{us1}, we considered the interactions of the Reggeon with SM gluons and quarks, but not their KK excitations.  There are two reasons why the latter can be important for Reggeon phenomenology.  First, both the Reggeon and the KK states are localised near the IR brane; the couplings are proportional to the overlap integrals and so should be sizeable.  Second, consider the mass spectrum of these states.  We show in figure~\ref{fig:MRegKK} the ratio of the Reggeon mass to the masses of various KK excitations as a function of the Reggeon five-dimensional mass parameter $m$.  These ratios are independent of $\LIR$, which serves only to set the overall scale.  For the KK gluon, this ratio depends only on $m$; for the KK quarks, it also depends on the $c$-parameters, and we have plotted a few representative values.
For all relevant values of $m$, the Reggeon is more massive than the lightest KK gluon and KK quarks.  For $m \gtrsim 2.2 k$, the Reggeon is more than twice as massive as the KK gluon and --- in model B --- the lighter KK quarks.  This means that decays of the Reggeon to an SM state and its KK partner are always kinematically allowed; decays to two KK excitations are allowed for larger $m$; and decays to two KK excitations with one off-shell may also be relevant for smaller $m$. Thus, a complete picture of the LHC phenomenology of the Reggeon must include its decays into final states containing the KK excitations of the SM particles. Let us consider these decays.

\begin{figure}[tb]
\begin{center}
\centerline {
\includegraphics[width=5.7in]{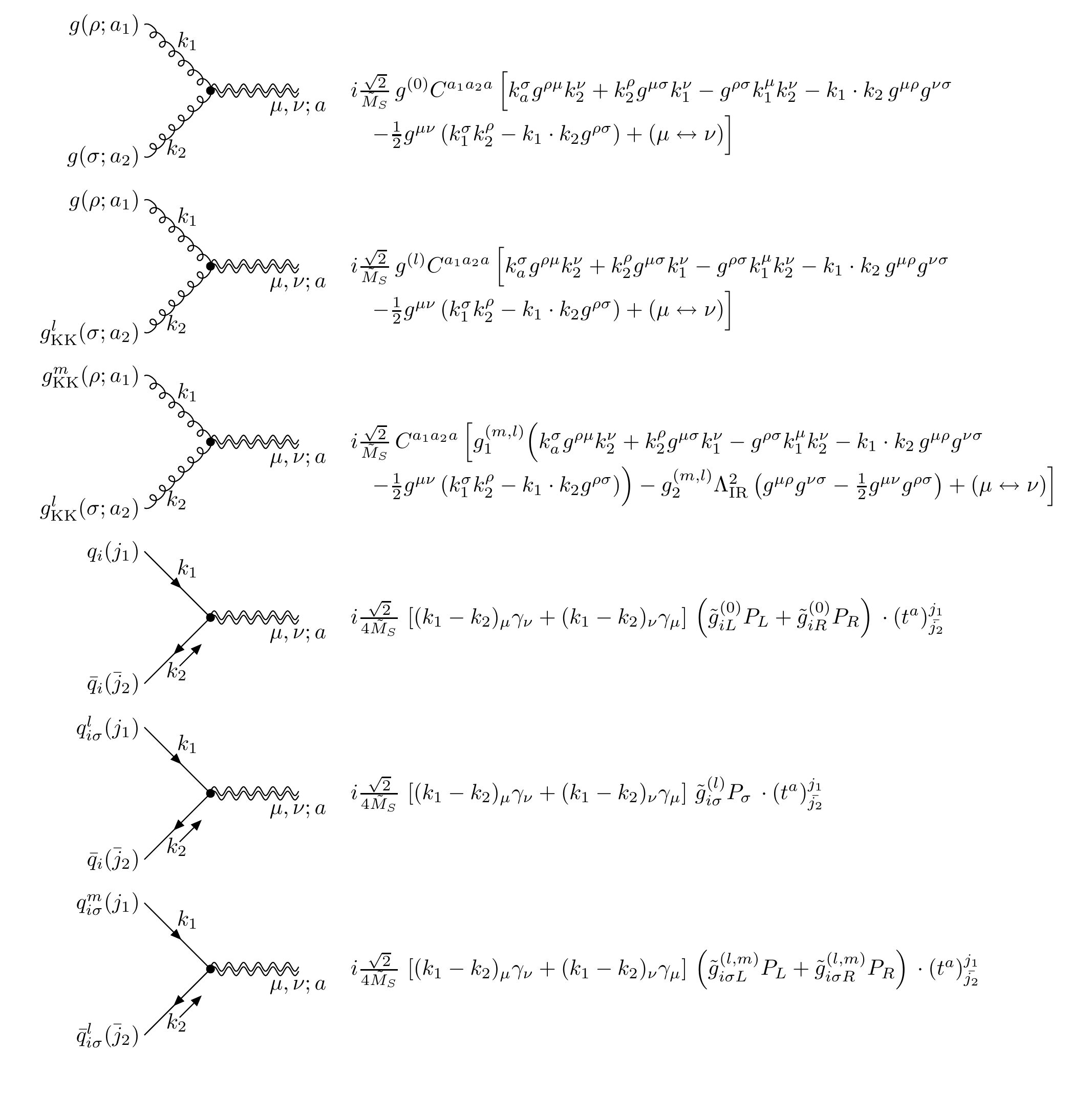}
}
\caption{The Feynman rules. Here the double wavy line denotes the lowest KK state of the Regge gluon $g^*$; $g$ denotes SM gluons; $g_{\rm KK}^l$ denotes the $l$-th level KK excitation of the SM gluon; $q_i$ denotes SM quarks, $i\in(u,d,s,c,b,t)$; and $q^l_{i\sigma}$ denotes the $l$-th level KK excitations of the quark with helicity $\sigma \in \{L, R\}$.}
\label{fig:Frules}
\end{center}
\end{figure}

The Feynman rules involving the spin-2 Reggeon, SM gluons and quarks, as well as their KK excitations, are listed in figure~\ref{fig:Frules}. Here we defined the warped-down string scale
\beq
\tilde{M}_S \,=\, e^{-\pi k R} M_S^*\,\sim\,{\rm a~few~TeV}.
\eeq{Mtilde}
The dimensionless coupling constants that enter the Reggeon-gluon vertices are given by
\beqa
g^{(0)} &=& \frac{g_s\,e^{-\pi k R}}{\pi R} \int_0^{\pi R} dy \, e^{2ky} \, f^{(0)} (y)\,,\CR%\hskip3cm
g^{(l)} &=& \frac{g_s\,e^{-\pi k R}}{\pi R} \int_0^{\pi R} dy \, e^{2ky} \, \chi^{(l)} (y)\,f^{(0)} (y)\,,\CR
g^{(l,m)}_1 & = & \frac{g_s\,e^{-\pi k R}}{\pi R} \int_0^{\pi R} dy \, e^{2ky} \, \chi^{(l)} (y)\, \chi^{(m)} (y)\, f^{(0)} (y)\,,\CR
g^{(l,m)}_2 & = & \frac{g_s\,e^{-\pi k R}}{\LIR^2 \pi R} \int_0^{\pi R} dy \, \chi^{(l)}{}' (y)\, \chi^{(m)}{}' (y)\, f^{(0)} (y)\,.
\eeqa{gluevertex}
The coupling constants entering the Reggeon-quark vertices are
\beqa
\tilde{g}_{i\sigma}^{(0)} &=& g_s \, e^{-\pi kR} k\, \frac{1-2c_{i\sigma}}{e^{\pi k R (1-2c_{i\sigma})} - 1}\, \int_0^{\pi R} \,dy\, e^{(3-2c_{i\sigma})ky}\, f^{(0)}(y)\,,\CR 
\tilde{g}_{i\sigma}^{(l)} &=& \frac{g_s \, e^{-\pi kR}}{\pi R} \sqrt{\frac{k \pi R (1-2c_{i\sigma})}{e^{\pi k R (1-2c_{i\sigma})} - 1}} \, \int_0^{\pi R} \,dy\, e^{(1-c_{i\sigma})ky}\, f^{(0)}(y) \, \varphi_{i\sigma\sigma}^{(l)} (y) \, ,\CR
\tilde{g}_{i\sigma_1\sigma_2}^{(l,m)} &=& \frac{g_s \, e^{-\pi kR}}{\pi R} \, \int_0^{\pi R} \,dy\, e^{- ky}\, f^{(0)}(y) \, \varphi_{i\sigma_1\sigma_2}^{(l)} (y) \, \varphi_{i\sigma_1\sigma_2}^{(m)*} (y) \, .  
\eeqa{quark_vertex}
As before, in these formulas $\sigma \in \{L, R\}$ and the index $i$ labels the SM quarks, $i\in (u,d,s,c,b,t)$.  
It is easy to estimate the size of the overlap integrals:
\begin{align}
  g^{(0)} & \sim \frac{1}{\sqrt{\pi k R}} ; & g^{(1)} & \sim 1 ; & g^{(1, 1)}_{1, 2} & \sim \sqrt{\pi k R} ; \notag \\
  \tilde{g}^{(0)}_{i\sigma} & \sim \frac{\sqrt{\pi k R} (1 - 2c_{i\sigma})}{1 - e^{(2c_{i\sigma} - 1) \pi k R}} ; & \tilde{g}^{(1)}_{i\sigma} & \sim \sqrt{\frac{\pi k R (1 - 2c_{i\sigma})}{1 - e^{(2c_{i\sigma} - 1) \pi k R}}} ; & \tilde{g}^{(1,1)}_{i\sigma_1\sigma_2} & \sim \sqrt{\pi k R} .
\end{align}
From eq.~\eqref{LIR_def}, $\pi k R \approx 35$.  This shows that the couplings involving light quarks are suppressed for $c > 1/2$ and enhanced for $c < 1/2$, as expected.  It also shows that the Reggeon couplings to the KK states are generally enhanced, by a factor of $\sim$ a few, compared to its couplings to the SM.

Using the Feynman rules in figure~\ref{fig:Frules}, it is straightforward to compute the partial decay widths of the Reggeon.  Introducing $\mu_G = m(g^1)/M$, $\gamma_G=\Gamma(g^1)/M$ and $\mu_{Fi\sigma} = m(f^1_{i\sigma})/M$, $\gamma_{Fi\sigma} = \Gamma(f^1_{i\sigma})/M$, we obtain 
\begin{subequations}\label{eq:RegPWid}
  \begin{align}
    \Gamma \bigl(g^\ast \to gg\bigr) & = \frac{\alpha_s M}{6} \biggl( \frac{g^{(0)} M}{g_s \tilde{M}_S} \biggr)^2 ; \\
    \Gamma \bigl(g^\ast \to gg^1\bigr) & = \frac{\alpha_s M}{3} \biggl( \frac{g^{(1)} M}{g_s \tilde{M}_S} \biggr)^2 \bigl( 1 - \mu_G^2 \bigr)^3 \bigl( 1 + \mu_G^2/2 + \mu_G^4/6 \bigr); \\
    \Gamma \bigl(g^\ast \to g^1g^{1(\ast)}\bigr) & = \frac{\alpha_s M}{6} \biggl( \frac{M}{\tilde{M}_S} \biggr)^2 %\left( h^2_G \bigl( g_1^{(1, 1)} , g_2^{(1, 1)}, \mu_G \bigr)\Theta(1-2\mu_G) + 
    h_G \bigl( g_1^{(1, 1)} , g_2^{(1, 1)}, \mu_G, \gamma_G \bigr) ;\\
    %\Theta(2\mu_G-1) \right); \\
    %%
    \Gamma \bigl(g^\ast \to f_{i\sigma}\bar{f}_{i\sigma}\bigr) & = \frac{\alpha_s M}{80} \biggl( \frac{\tilde{g}_{i\sigma}^{(0)} M}{g_s \tilde{M}_S} \biggr)^2 ; \\
    \Gamma \bigl(g^\ast \to f_{i\sigma}\bar{f}_{i\sigma}^1\bigr) & = \Gamma \bigl(g^\ast \to f^1_{i\sigma}\bar{f}_{i\sigma}\bigr) = \frac{\alpha_s M}{80} \biggl( \frac{\tilde{g}_{i\sigma}^{(0)} M}{g_s \tilde{M}_S} \biggr)^2 \bigl( 1 - \mu_{Fi\sigma}^2 \bigr)^4 \bigl( 1 + 2\mu_{Fi\sigma}^2/3 \bigr) ;\\
    \Gamma \bigl(g^\ast \to f^1_{i\sigma}\bar{f}^{1(\ast)}_{i\sigma}\bigr) & = \frac{\alpha_s M}{80} \biggl( \frac{M}{\tilde{M}_S} \biggr)^2 
    %\left(h^2_G \bigl( \tilde{g}^{(1, 1)}_{i\sigma L}, \tilde{g}^{(1, 1)}_{i\sigma R}, \mu_{Fi\sigma}) 
%\Theta(1-2\mu_{Fi\sigma})  +  
    h_F \bigl( \tilde{g}^{(1, 1)}_{i\sigma L}, \tilde{g}^{(1, 1)}_{i\sigma R}, \mu_{Fi\sigma}, \gamma_{Fi\sigma} \bigr)  \,.
    %\Theta(2\mu_{Fi\sigma}-1)\,.
    %\right).
  \end{align}
\end{subequations}
We have assumed that we can treat all SM particles, including top quarks, as massless.  The decays to two KK excitations include both the above-threshold two-body final states, and the below-threshold three-body final states.  The functions $h_G$ and $h_F$ are given in appendix~\ref{3body}.

For the KK gluon, the dominant decay is to $t\bar{t}$~\cite{KKgluon} and $\Gamma(g^1) \approx 0.153 \, m(g^1)$. (This formula applies in both models A and B, since the top wavefunctions are identical.) The dominant decay of KK quarks is to $Wq$ or $Zq$~\footnote{Decays to $g^1q$ are enhanced relative to electroweak channels by $g_s \sqrt{\pi k R}/g_w$, but suppressed by phase space and $m_{W,Z}^2/m(g^1)^2$.}. A precise estimate of the KK quark width depends on the flavour structure of the model.  As a conservative estimate, we assume that they decay to $W q$ with an effective coupling $g \approx 1$. For the most phenomenologically relevant case, the KK partners of the first two generations in Model B, this should be an overestimate of the actual coupling strength, and thus of the Reggeon partial width in this channel below threshold. Even so, as we will see below, the predicted branching ratio of this channel is negligible, so a more precise estimate is unnecessary. 

\begin{figure}[tb]
\begin{center}
\centerline {
\includegraphics[width=4in]
{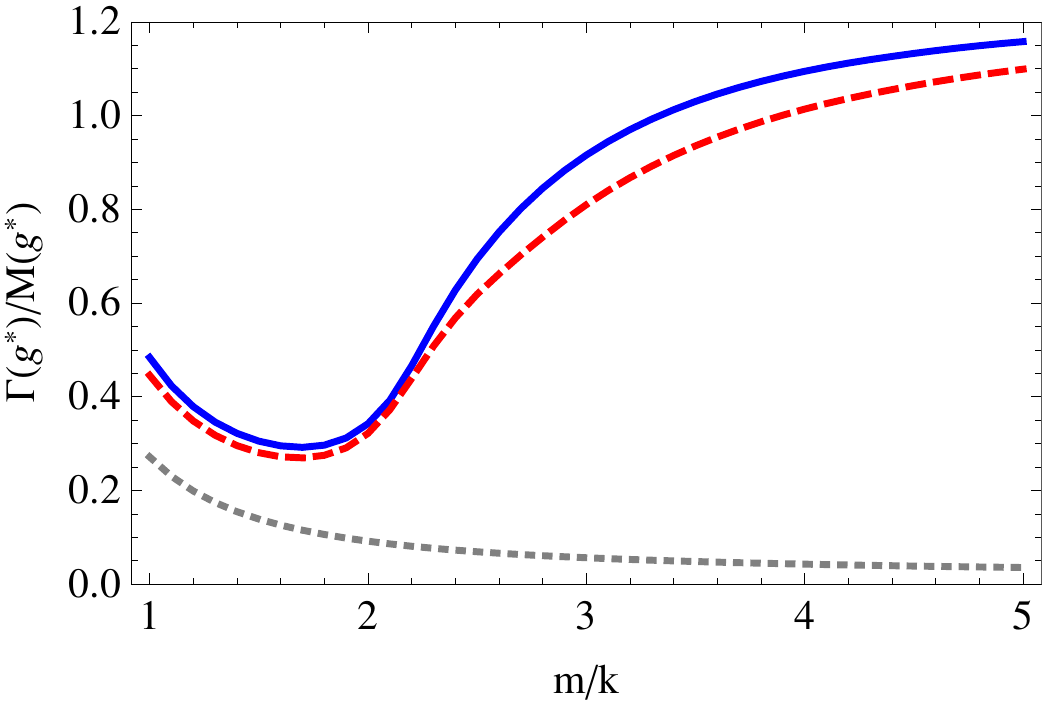}
}
\vspace{-0.2in}
\caption{The Reggeon total decay width as a function of $m/k$ in Model A (red/dashed line) and Model B (blue/solid line). The gray/dotted line shows the partial width of direct decays to SM states such as $t\bar{t}$.}
\label{fig:Width}
\end{center}
\end{figure}

In figure~\ref{fig:Width}, we plot the total Reggeon width, as a fraction of its mass, for the two sets of $c_i$'s given in eqs.~\leqn{cs_Higgs},~\leqn{cs_HL}. Note that the quantity $\Gamma(g^*)/M(g^*)$ is to a good approximation independent of $\LIR$, which sets the overall mass scale for both the Reggeon and the KK states, and depends only on the ratio $m/k$ or, equivalently (see figure~\ref{fig:Rmass}), $M/\LIR$. 
We also show the Reggeon width to SM states only, computed in ref.~\cite{us1}. It is obvious that the extra decay channels significantly increase the Reggeon width.  Once decays to two on-shell KK states become available, the predicted width of the Reggeon increases dramatically, eventually becoming of the same order as its mass.  This places a limitation on our model: by construction, we treated the Reggeon as a narrow particle which can appear as an asymptotic state in the $S$ matrix. We therefore restrict our attention to Reggeons below this threshold.

\begin{figure}[tb]
\begin{center}
\centerline {
\includegraphics[width=3in]
{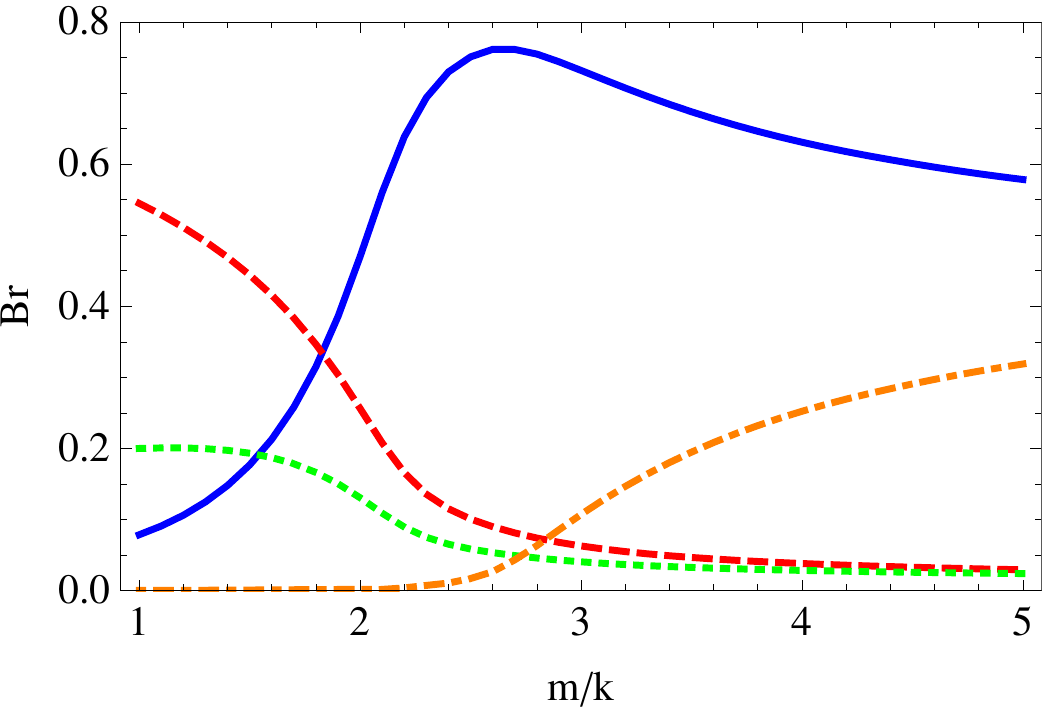}
\includegraphics[width=3in]
{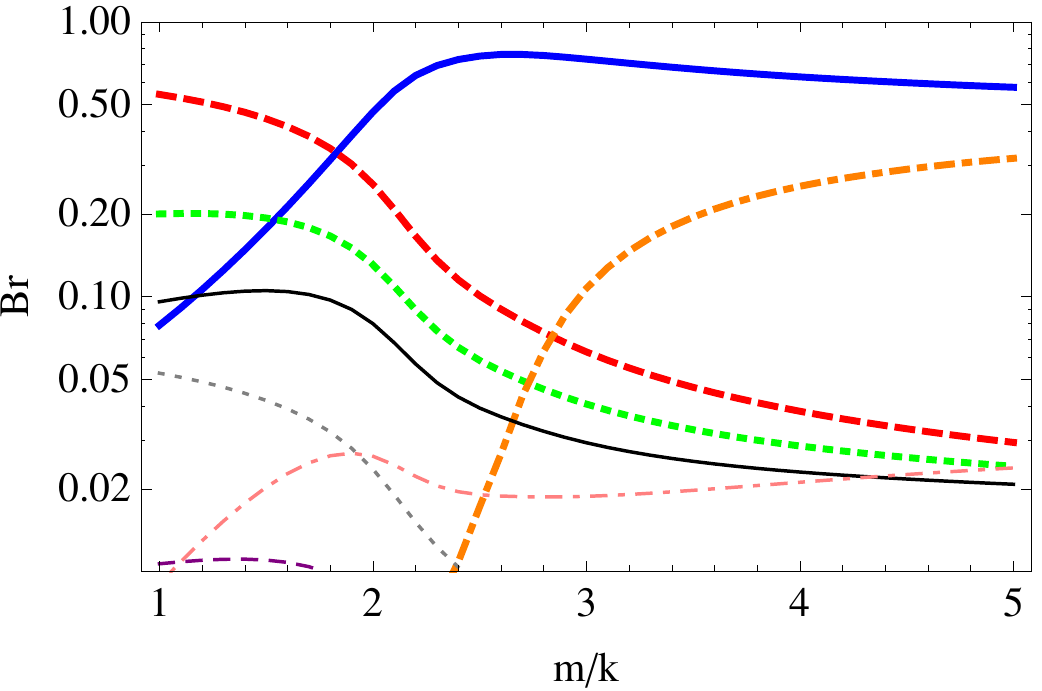}
}
\caption{The Reggeon branching fractions in Model A: (left) The four leading decay channels; (right) All channels with branching ratio above 1\%. On the left panel, the blue solid line corresponds to the $g^1 g^{1(\ast)}$ final state; the red dashed line to the $t_R \bar{t}_R$; the green dotted line to $g^1g$; and the orange dot-dashed line to two KK quarks (all flavors). The additional thin lines on the right panel are: 
$t_L \bar{t}_L^1 + b_L \bar{b}_L^1 + t_L^1 \bar{t}_L + b_L^1 \bar{b}_L$ (solid); quark + KK quark summed over first two generations + $b_R$ (dashed); $t_L \bar{t}_L + b_L \bar{b}_L$ (dotted); and $t_R \bar{t}_R^1 + t_R^1 \bar{t}_R$ (dot-dashed).}
\label{fig:BR_A}
\end{center}
\end{figure}

\begin{figure}[h]
\begin{center}
\centerline {
\includegraphics[width=3in]
{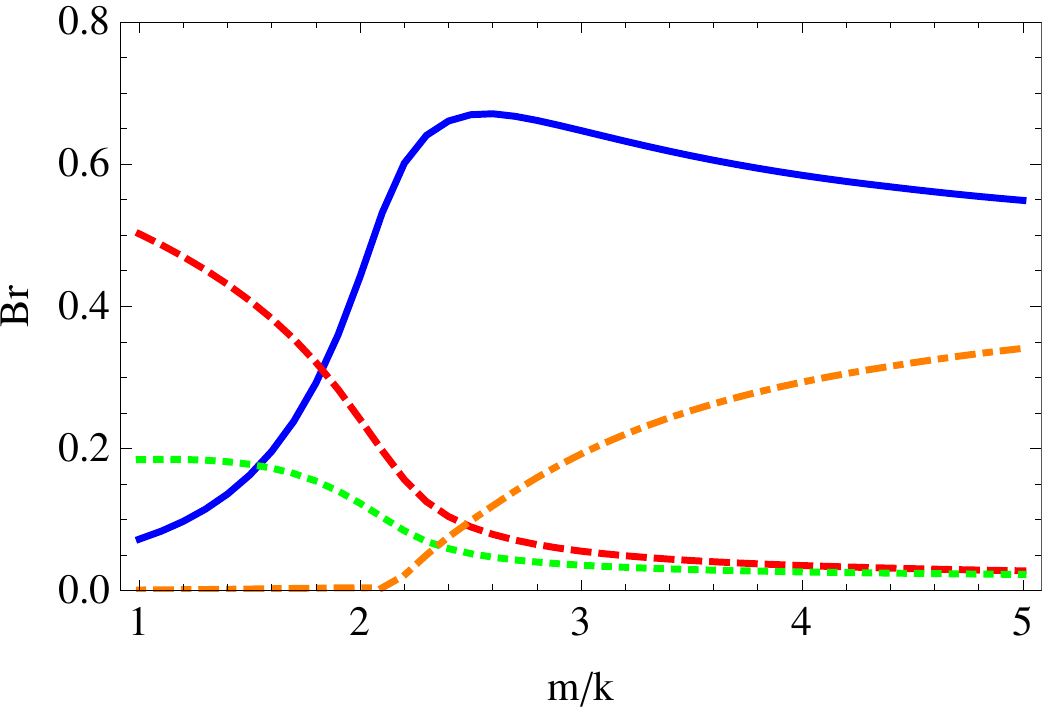}
\includegraphics[width=3in]
{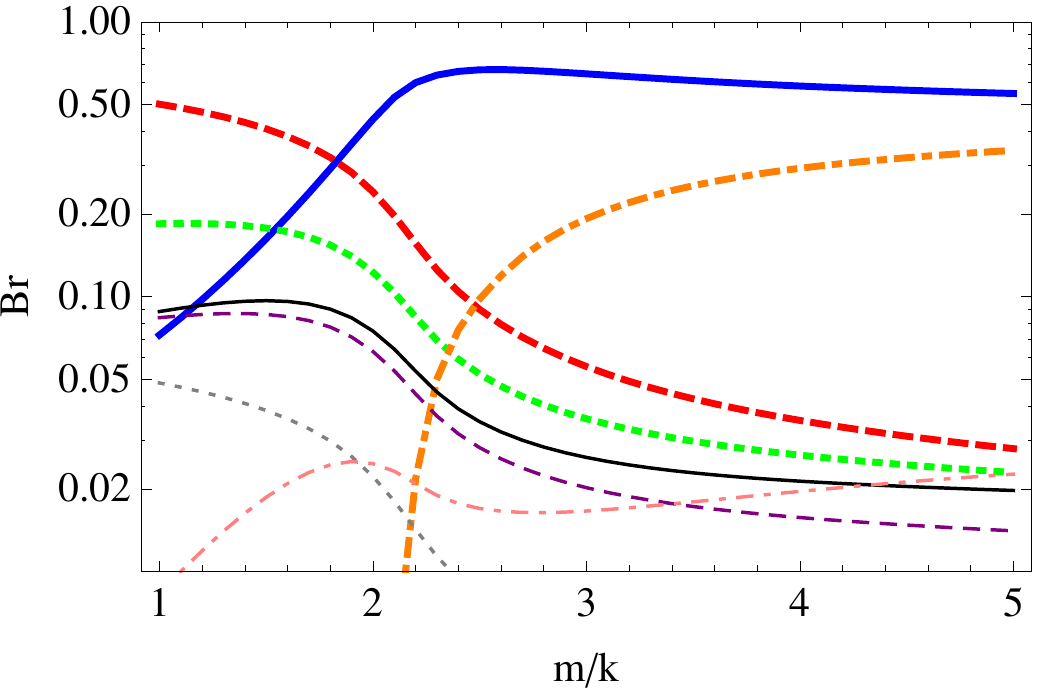}
}
\caption{The Reggeon branching fractions in Model B: (left) The four leading decay channels; (right) All channels with branching ratio above 1\%. The color scheme is the same as in figure~\ref{fig:BR_A}.}
\label{fig:BR_B}
\end{center}
\end{figure}

In figures~\ref{fig:BR_A} and~\ref{fig:BR_B}, we plot the Reggeon branching fractions for the two models.
Again, the branching fractions are independent of $\LIR$ and depend only on $m/k$ or $M/\LIR$.
It is immediately clear that decays involving KK excitations of the SM play a very important role. The largest of the direct decays to SM zero-modes, $g^*\to t\bar{t}$, dominates only for very light Reggeons, $m/k \approx 1$.  
In most of the parameter space in both models, the decay $g^\ast \to g^1g^1$ is dominant.  As noted, the KK gluon decays with almost 100\% probability to $t\bar{t}$. Since its mass is well above $m_t$, the tops from this decay are typically relativistic, resulting in ``top jets" in the detector.  The resultant LHC signal --- four boosted tops with nearly equal energies --- is quite dramatic, and plausibly visible despite the broad nature of the resonance. 

\section{Four boosted top signature at the LHC}
\label{sec:LHC}

In this section, we present a preliminary study of the observability of the four-top signature of the Reggeon at the LHC. Throughout this section, we fix the 5D Reggeon mass parameter to be $m=2k$, and vary the physical 4D Reggeon mass by changing $\LIR$. With this choice, the Reggeon branching ratios and other phenomenologically important dimensionless quantities remain fixed, independent of its mass, simplifying the analysis. In particular, the ratio of the Reggeon mass to that of the first KK excitation of the gluon is about 1.9; the branching ratio of the decay $g^*\to g^1 t \bar{t}$, which produces the 4-top signature we're interested in, is 47\% in Model A and 44\% in Model B; and the total width of $g^*$, in units of its mass, is about 0.35, so that narrow-width approximation can be trusted.
Of course, for a more thorough understanding of the LHC reach in this channel one should explore varying both $\LIR$ and $m/k$; such an analysis is outside the scope of this paper.

 \begin{figure}[tb]
\begin{center}
\centerline {
\includegraphics[width=3in]
{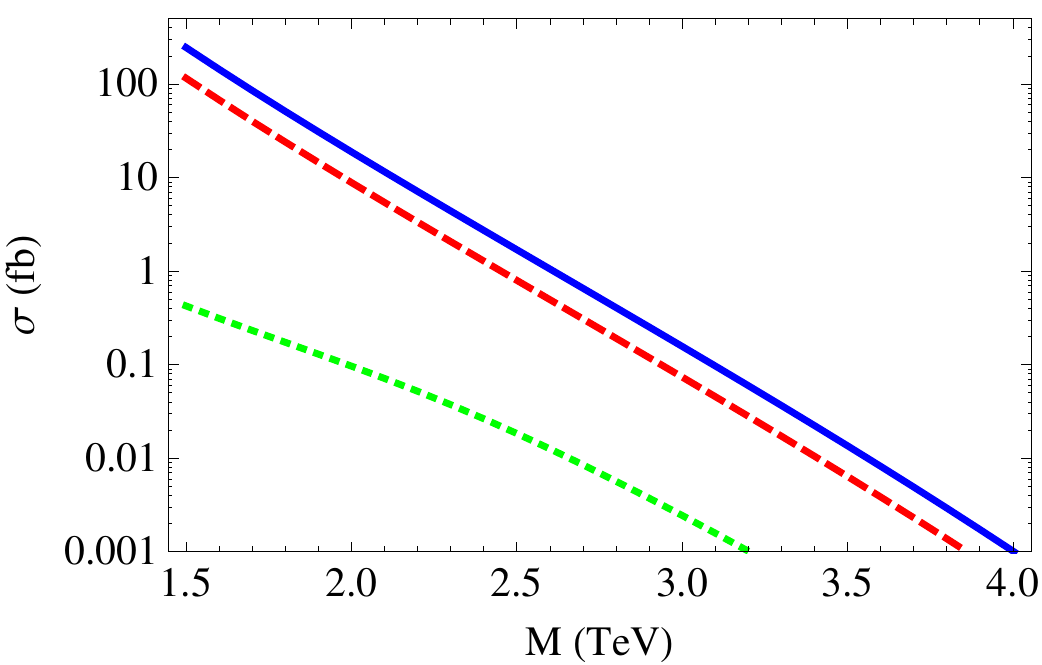}
\includegraphics[width=3in]
{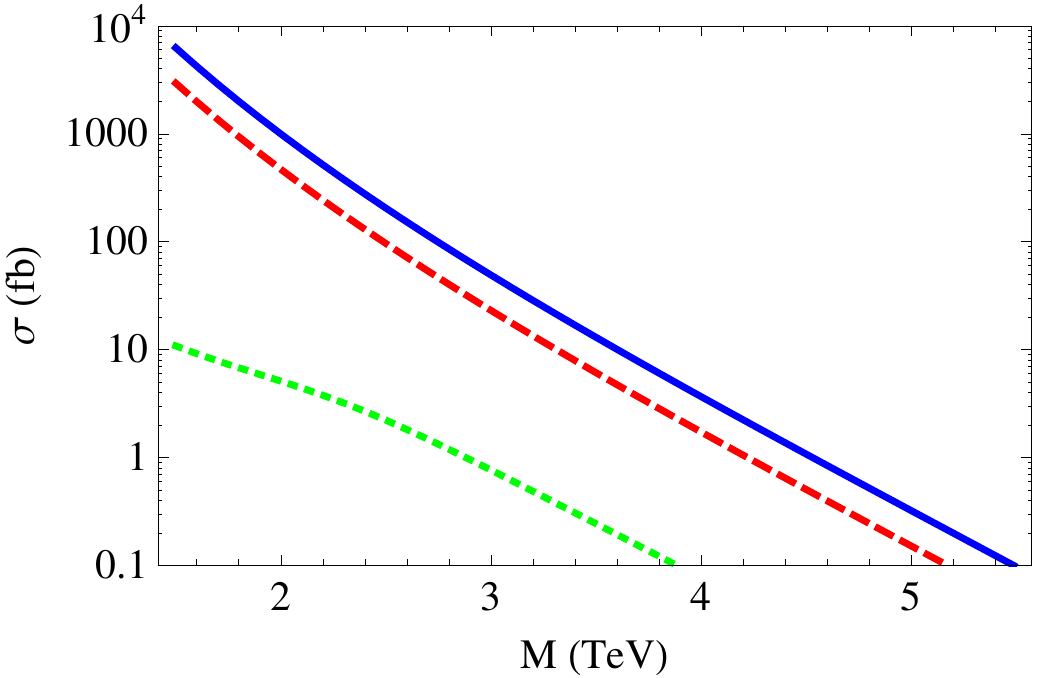}
}
\caption{The Reggeon production cross section, as a function of its mass, in Model A: (left) $\sqrt{s}=7$~TeV; (right) $\sqrt{s}=14$~TeV.  We used the MSTW 2008~\cite{mstw2008} PDF set at next to leading order, with the factorization and renormalization scales set to the Reggeon mass. %\draftnote{Specify PDF set used and factorization/renormalization scale.} 
In both panels, blue/solid line corresponds to the total production cross section; red/dashed lines show the total rate of the four-top events; and green/dotted lines show the rate of events for which all four top-jets are tagged. }
\label{fig:xsec_A}
\end{center}
\end{figure}

\begin{figure}[h]
\begin{center}
\centerline {
\includegraphics[width=3in]
{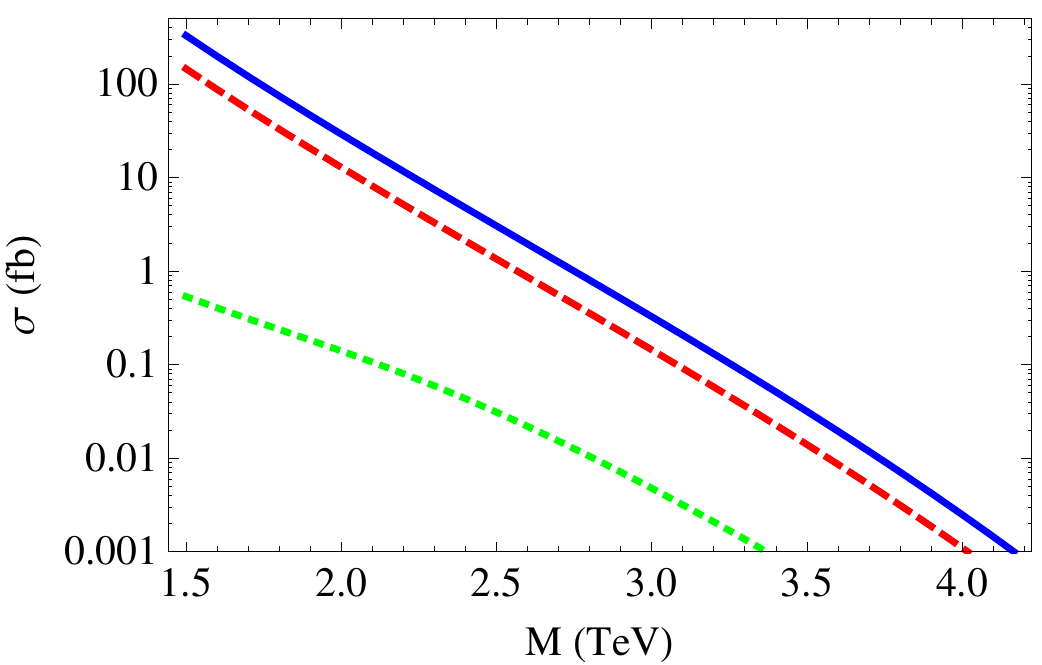}
\includegraphics[width=3in]
{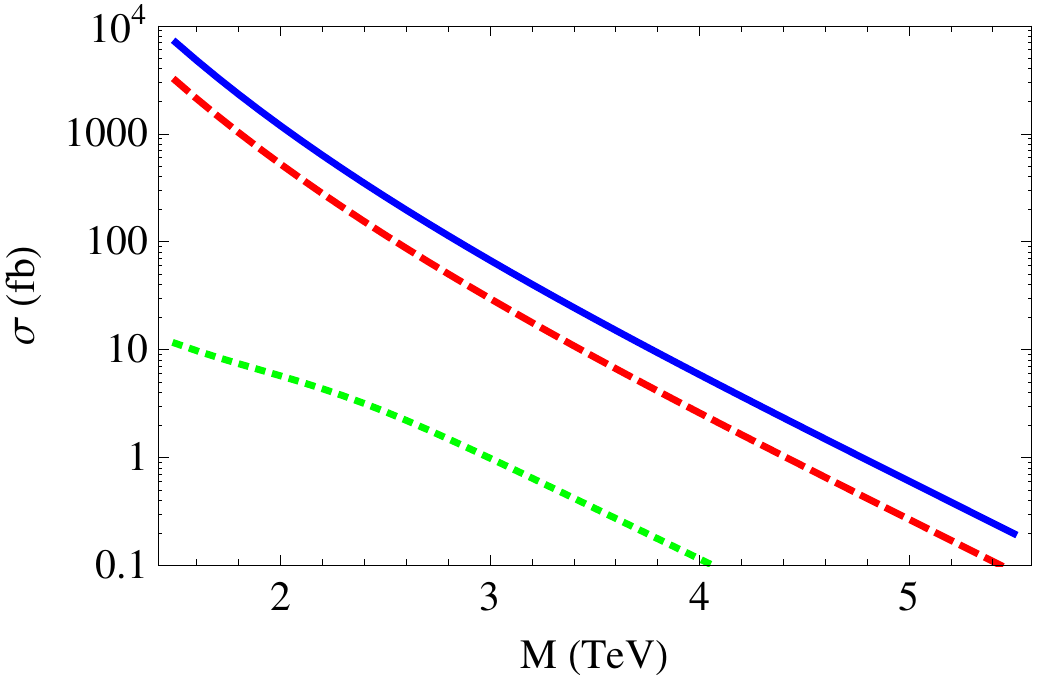}
}
\caption{The Reggeon production cross section, as a function of its mass, in Model B: (left) $\sqrt{s}=7$ TeV; (right) $\sqrt{s}=14$ TeV. The notation is the same as in figure~\ref{fig:xsec_A}.}
\label{fig:xsec_B}
\end{center}
\end{figure}

In much of the interesting parameter space of the model, the Reggeons are expected to be copiously produced at the LHC. The total production cross sections at $\sqrt{s}=7$~TeV and $\sqrt{s}=14$~TeV, computed using the leading-order parton-level cross sections given in~\cite{us1}, are shown by the blue/solid lines in figure~\ref{fig:xsec_A} and figure~\ref{fig:xsec_B}.\footnote{Note that the total cross sections at 14 TeV shown here do not correspond to fig.~3 of ref.~\cite{us1}, due to different choices of $\LIR$ and $m/k$ made in that paper.}  The red/dashed lines in the same figures show the total rate of the four-top events from Reggeon decays. For the Reggeon mass around 2 TeV,\footnote{We note that a 2 TeV Reggeon corresponds to $\LIR\approx$ 0.5 TeV and a KK gluon at approximately 1 TeV, which is consistent with precision electroweak constraints in Model B though not in model A. In either model, such low values of $\LIR$ are disfavored by flavor constraints. Still, a direct search for the Reggeon in this parameter range is very worthwhile, since electroweak and flavor constraints are indirect and may be evaded, for example, due to additional, model-dependent new physics contributions.} we expect $O(100)$ events 
to appear in a 10 fb$^{-1}$ sample at 7 TeV center-of-mass energy, a conservative estimate for what may be collected in the current 2011-12 LHC run. At 14 TeV collision energy and design luminosity, significant samples of Reggeons should be available for Reggeon masses up to about 5 TeV.

Since the KK gluon mass is close to half of the Reggeon mass, the four tops produced in the process $g^*\to g^1 t\bar{t}$, followed by $g^1\to t\bar{t}$, have roughly equal energies in the Reggeon rest frame.  The high mass of the Reggeon implies that it is produced approximately at rest in the lab frame, so we can estimate the energy of each top in the lab frame to be approximately $M/4$.  In the interesting parameter range, this energy is above 400 GeV, meaning that the tops are boosted and their decay products will be collimated, so that each top will likely be identified as a single jet in the detector. Experimental discrimination between such ``top-jets" and ordinary QCD jets initiated by non-top quarks or gluons has been the subject of much recent work. A useful recent summary of the status of the field is given in ref.~\cite{boost2010}, which describes several algorithms designed for this purpose.  These ``top-taggers'' can be characterized by an efficiency and a fake rate, both of which are $p_T$-dependent.  We use the CMS tagger described in ref.~\cite{boost2010}. It has a maximal fake rate of 5\%,
no sensitivity for jets with $p_T < 250$~GeV and achieves maximal efficiency of 50\%
 for $p_T > 600$~GeV.  For jets between these limits, we model the $p_T$-dependence of the efficiency as linear, which understates the true efficiency.  As the Reggeon and KK gluons are approximately at rest and we have averaged over spins, the angular distribution of tops from each KK gluon is approximately isotropic in the lab frame. Within this approximation, we can determine the $p_T$ distribution of the tops and thus estimate the signal efficiency to tag the jets as tops.  The efficiency is a function of the Reggeon mass: for low Reggeon masses, the tops have lower $p_T$ and so the efficiency drops. 
 %for example, for a 2~TeV Reggeon it is about 1.1\%. 
The green/dotted lines in figures~\ref{fig:xsec_A},~\ref{fig:xsec_B} show the expected rate of events with  four jets tagged as tops.  Of course, it may not be necessary to demand that all four jets be tagged: doing so yields the best possible signal/background ratio, but may result in a loss of statistics. We will not attempt to optimize the search in this paper; our goal is simply to demonstrate that at least some of the interesting 
parameter space can be covered. 
  
 \begin{table}[t!]
		\renewcommand{\arraystretch}{1.2}
\begin{center}
\begin{tabular}{|l||r|r|r|r|} \hline
process  & $\sigma_{\rm tot}$ & Prob(4 top-tags)  & Eff($p_T>250$ GeV) &  $\sigma_{\rm tot} \cdot Prob \cdot Eff $
\rule{0ex}{2.2ex} \\ \hline \hline
signal     & 147                & $3.66 \times 10^{-3}$ &  & 0.54 \\ \hline\hline
$4j$       & $5.16 \times 10^{5}$ & $6.25 \times 10^{-6}$ & $7.0 \times 10^{-4}$ & $2.3 \times 10^{-3}$ \\ \hline
$3j+t$     & $1.35 \times 10^{5}$ & $6.25 \times 10^{-5}$ & $1.0 \times 10^{-4}$ & $8.4 \times 10^{-4}$ \\ \hline
$2j+2t$    & $1.63 \times 10^{3}$ & $6.25 \times 10^{-4}$ & $4.2 \times 10^{-3}$ & $4.3 \times 10^{-3}$ \\ \hline
$1j+3t$    & 0.221                & $6.25 \times 10^{-3}$ & $6.8 \times 10^{-3}$ & $9.4 \times 10^{-6}$ \\ \hline
$4t$       & 0.442                & 0.0625                & $7.7 \times 10^{-3}$ & $2.1 \times 10^{-4}$ \\ \hline\hline
Total Bg   &&&& 7.6 $\times 10^{-3}$ \\ \hline
\end{tabular} \\[1ex]
\caption{Signal and background cross sections (in fb), before and after cuts, at $\sqrt{s}=7$ TeV.  The signal is for a 2~TeV Reggeon in Model B.}
\label{tab:SBg7}
\end{center}
	\renewcommand{\arraystretch}{1.}
\end{table}
 
\begin{figure}[t]
\begin{center}
\centerline {
\includegraphics[width=4in]
{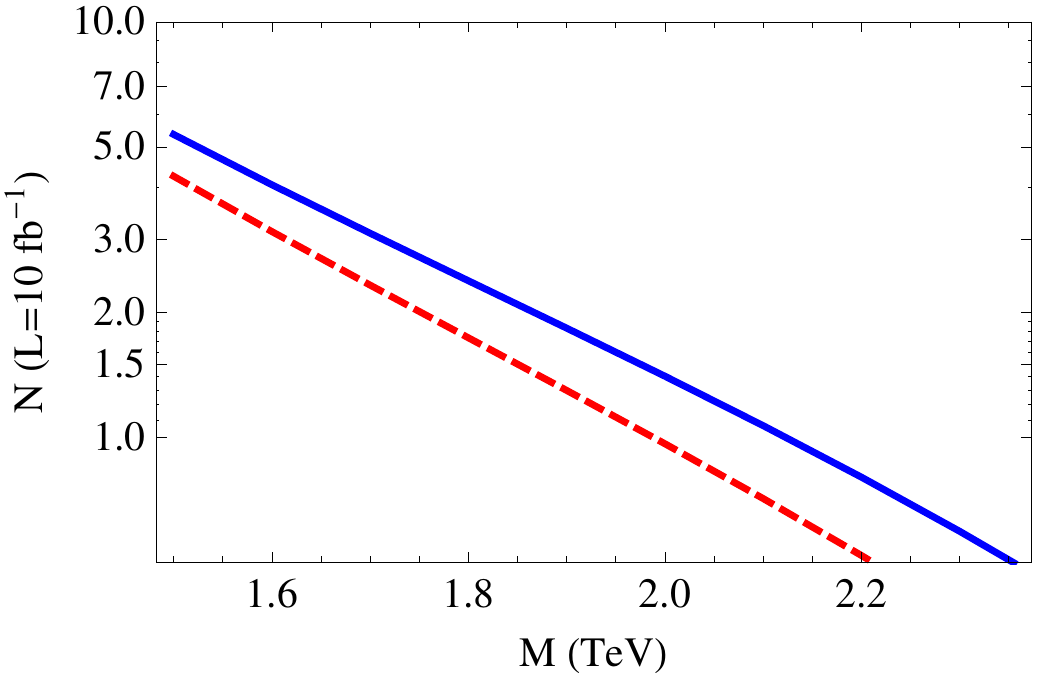}
}
\caption{Expected number of events in a Reggeon search in the four-top channel at the 7 TeV LHC with 10 fb$^{-1}$ integrated luminosity, in model A (red/dashed lines) and model B (blue/solid lines).} %Thick lines indicate the predicted number of signal events (including cut efficiencies), and thin lines indicate statistical significance, in $\sigma$, of the excess over the SM prediction of about 0.1 event.}
\label{fig:disc7}
\end{center}
\end{figure} 
 
The irreducible backgrounds consist of SM processes producing $n$ tops and $4-n$ QCD jets, with $n=0\ldots 4$. We studied these backgrounds at the parton level, using the {\tt MadGraph/MadEvent v4} 
Monte Carlo event generator~\cite{MGME}. The background rates for $\sqrt{s}=7$ TeV are listed in table~\ref{tab:SBg7}. To render the signal observable, we demand 4 top-tagged jets with $p_T>250$ GeV.  For simplicity and to be conservative, we did not model the tagging efficiency in the background calculation beyond applying the $p_T$ cut; rather, we imposed a constant 50\% efficiency and 5\% fake rate.
The total rate of background events satisfying these requirements is only about 0.01 fb, so that the search in this channel is essentially background-free in the current run. For comparison, we also list the signal rate, corresponding to the Reggeon of $M=1.5$~TeV in Model B; in this case the tagging efficiency includes an estimate of the effect of $p_T$ dependence, as described above. About 5 events are expected in a 10 fb$^{-1}$ data set, compared to expected background of 0.1 event. We conclude that a 1.5~TeV Reggeon of our model can be easily seen, or convincingly ruled out, by the current LHC run.  The number of expected events in the current run is shown in figure~\ref{fig:disc7}.  In model A (model B), we can potentially exclude at 95\%
Bayesian credibility Reggeons with masses up to 1.6~(1.7)~TeV; and weak evidence --- at least one event --- is expected up to 2~(2.1)~TeV.

\begin{table}[t!]
		\renewcommand{\arraystretch}{1.2}
\begin{center}
\begin{tabular}{|l||r|r|r|r|} \hline
process  & $\sigma_{\rm tot}$ & Prob(4 top-tags)  & Eff($p_T>250$) &  $\sigma_{\rm tot} \cdot Prob \cdot Eff $
\rule{0ex}{2.2ex} \\ \hline \hline
signal     & 6.12                & 0.0398 & & 0.24 \\ \hline\hline
$4j$       & $1.96 \times 10^{10}$ & $6.25 \times 10^{-6}$ & $9.1 \times 10^{-7}$ & $1.1 \times 10^{-1}$ \\ \hline
$3j+t$     & $6.18 \times 10^5$    & $6.25 \times 10^{-5}$ & $4.0 \times 10^{-5}$ & $1.6 \times 10^{-3}$ \\ \hline
$2j+2t$    & $4.34 \times 10^5$    & $6.25 \times 10^{-4}$ & $4.0 \times 10^{-4}$ & $1.1 \times 10^{-1}$ \\ \hline
$1j+3t$    & 0.137                 & $6.25 \times 10^{-3}$ & $3.1 \times 10^{-2}$ & $3.3 \times 10^{-5}$ \\ \hline
$4t$       & 9.72                  & 0.0625                & $2.3 \times 10^{-2}$ & $1.4 \times 10^{-2}$ \\ \hline\hline
Total Bg   & 1.96 $\times 10^{10}$ &&& 0.24 \\ \hline
\end{tabular} \\[1ex]
\caption{Signal and background cross sections (in fb), before and after cuts, at $\sqrt{s} = 14$~TeV.  The signal is for a 3.5~TeV Reggeon in Model A.}
\label{tab:SBg14}
\end{center}
	\renewcommand{\arraystretch}{1.}
\end{table}

\begin{figure}[h]
\begin{center}
\centerline {
\includegraphics[width=4in]
{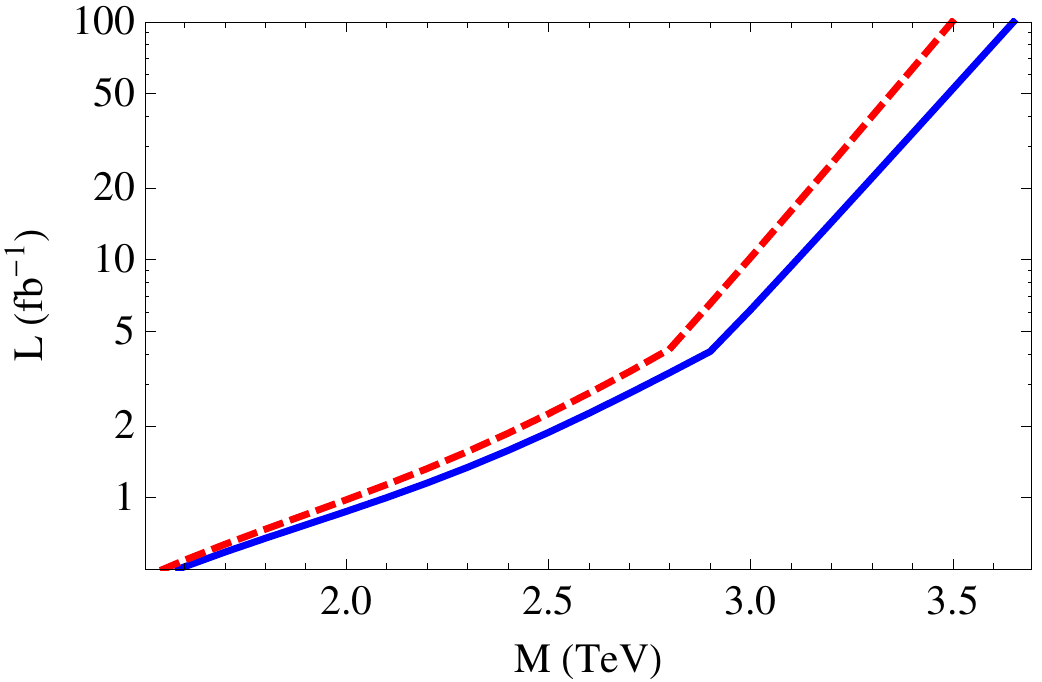}
}
\caption{Estimated luminosity needed to discover the Reggeon at a $5\sigma$ level (statistics-only) and requiring at least 5 events, at the 14 TeV LHC. The notation is the same as in figure~\ref{fig:disc7}.}
\label{fig:disc14}
\end{center}
\end{figure}

The estimated background rates for the LHC run at the design energy, 14 TeV, are listed in table~\ref{tab:SBg14}.  We also show the signal for a 3.5~TeV Reggeon in model A, for which the signal and backgrounds are equal.  (The equivalent point for model B is 3.7~TeV.)  In figure~\ref{fig:disc14}, we plot the integrated luminosity needed for $5\sigma$ discovery of the Reggeon, ignoring systematic errors.  We also demand that at least 5 events be observed; this limits discovery for Reggeons below 3~TeV where 5$\sigma$ statistical significance is easily acheived. With realistic luminosities, much of the interesting parameter range would be covered by this search.  For heavier Reggeons, we might benefit from demanding fewer jets tagged as tops and properly accounting for the $p_T$-dependence of the fake rate. More detailed studies are needed to determine the ultimate reach of the LHC in this channel.

\section{Current Experimental Constraints}

It is worth commenting briefly on current experimental limits on this model.  Our philosophy is to ignore indirect limits, as those will depend on model-dependent new physics contributions.  In particular, the stringy contributions to flavour and electroweak precision constraints can not be calculated within our simplified model. We can potentially find limits using direct searches for dijet, multijet and ditop resonances at the Tevatron and the LHC.  For the Reggeon itself, most such searches have questionable validity due to the large width.  We would not expect standard bump-hunting algorithms to find the Reggeon.  However, even if we ignore this point the Reggeon easily evades the current limits; the strongest constraints come from searches for dijet resonances at the LHC~\cite{Collaboration:2011ns}.  These searches currently limit the cross section times branching ratio at 1.5~TeV to be 200~fb; this should be compared to $\lesssim$~100~fb for the Reggeon.  (At higher masses, our signal cross section decreases more rapidly than the experimental limits.)  Limits on ditop searches are weaker, with upper limits on $\sigma \times B$ of $\sim$~0.5~pb in the relevant mass range~\cite{CMS-PAS-EXO-11-006}.

We can also consider limits on the KK gluon and quarks.  These objects are lighter and narrower than the Reggeon, and so one would expect them to be more strongly constrained; however, they also couple less efficiently to the gluon.\footnote{The gluon-gluon-KK gluon overlap integral vanishes, by orthogonality; the same is also true for the gluon-gluon-KK quark overlap integrals for the lightest KK quarks, the first two generations in Model B.}  The current LHC limits on the KK gluon are 650~GeV~\cite{ATLAS-CONF-2011-087}; the limits for KK quarks would be expected to be weaker, due to smaller cross sections and less distinctive decays.

Finally, one might be concerned about spin-3/2 Regge excitations of the quarks, which would have masses comparable to the Reggeon~\cite{Regge_top}.  It is difficult to use limits on Regge quarks to constrain our model, as the Regge quark mass and single production cross section both depend on new parameters beyond our framework.  The specific framework of~\cite{Regge_top} finds that single production of TeV-scale Regge quarks has a cross section of 10s of fb, with dominant decay mode to top + jet.  This lies below the dijet search limits already quoted, so that no further constraints can be inferred from these objects.

In summary, we conclude that the entire model parameter space where the four boosted top signature discussed in section~\ref{sec:LHC} is useful is currently viable.

\section{Conclusions and outlook}
\label{sec:conc}

In this paper, we continued our study of the LHC phenomenology of the Regge excitations of the SM particles in the framework of RS models with all matter in the bulk. We used a field-theoretic toy model of the tensor (spin-2) Regge excitation of the gluon, which was constructed in ref.~\cite{us1}. We extended the phenomenological analysis of~\cite{us1} to include the decays of the Regge gluon to final states containing Kaluza-Klein (KK) excitations of SM particles. We found that these decays play an important role in the Reggeon phenomenology. In particular, the decay to two KK excitations of the gluon (one of which may be off-shell) was found to have a large branching fraction in much of the interesting parameter space. This decay leads to a final state containing four top quarks. Typically, these top quarks are boosted (relativistic) in the lab frame, resulting in four ``top jets" in the detector, a highly distinctive signature. With realistic assumptions about top-tagger efficiencies and fake rates based on recent studies in the literature, we estimate that the LHC search for the Reggeons in this channel could cover some of the interesting parameter space already in the current 7 TeV run, and probe the Reggeon masses up to at least about 3.5 TeV in the future 14 TeV run. The results are not strongly dependent on the details of the model, such as the wavefunctions of SM fermions in the 5th dimension.

The obvious next step is to incorporate the Reggeon into a Monte Carlo generator. A variety of factors that could affect the efficiency of the search, such as kinematic distributions of the tops produced in Reggeon decays, initial-state radiation, showering and hadronization, etc., can be easily analyzed once this is achieved, improving the estimates of this paper. Of course, Monte Carlo samples of events with Reggeon production and decay are also necessary to compare the predictions of the model with the LHC data. Once the signal Monte Carlo is available, searches for the Reggeon should be undertaken by the LHC collaborations.  

\vskip1cm

%\noindent{\large \bf Acknowledgments} 
\acknowledgments

We thank Monika Blanke,  David Morrissey, and Jing Shu for useful discussions. MP is supported by the U.S. National Science Foundation through grant PHY-0757868 and CAREER award PHY-0844667. APS is partially supported by the Natural Science and Engineering Council of Canada.

\begin{appendix}

\section{Three-body decay widths of the Reggeon}
\label{3body}

In this Appendix, we present the explicit form of the functions $h_G$ and $h_F$ which enter the three-body decay widths of the Reggeon to $g^1 t\bar{t}$ and $q^1 q W$, respectively. The function $h_G$ is given by
\beqa
h_G \bigl( g_1 , g_2, \mu, \gamma \bigr) &=&    \frac{1}{12\pi} \frac{\gamma}{\mu^3} \int_{2\mu}^{1+\mu^2} \ud x \frac{\sqrt{x^2 - 4\mu^2}}{(1 - x)^2 + \mu^2 \gamma^2} \CR
  & & \times \biggl[ \biggl( \frac{g_1}{g_s} \biggr)^2 2 \mu^2 \Bigl\{ - x^5 + x^4 (6 + \mu^2) - x^3 (15 + 7\mu^2) + 2x^2 (5 + 26\mu^2 + \mu^4) \CR
  & & \qquad\qquad\qquad - 4x\mu^2 (15+19\mu^2) + 4\mu^2 (5 + 14 \mu^2 + 9\mu^4) \Bigr\} \CR
 & & + \biggl( \frac{g_1 g_2 \LIR^2}{g_s^2 M^2} \biggr) 40 \mu^2 \Bigl\{ x^3 - x^2 (4 + \mu^2 ) + x (3 + 5\mu^2 ) - 2 \mu^2 - 2\mu^4 \Bigr\} \CR
& &  + \biggl( \frac{g_2 \LIR^2}{g_s M^2} \biggr)^2 \Bigl\{ x^4 - 10x^3 + 2x^2 (5 + 6\mu^2) - 80 x\mu^2 + 8\mu^2 (10+7\mu^2) \Bigr\} \biggr] \,.\CR
\eeqa{3bodyG}
It is worth noting that this expression contains an additional factor of $1/2$ compared to a naive calculation of the three-body width.  This is a phase space factor, that can most easily be understood by considering the four-body width $g^\ast \to g^1g^1 \to t\bar{t}t\bar{t}$.  This process has two diagrams and a phase space symmetry factor of $1/4$ (two pairs of identical particles in the final state), so that when we make the narrow width approximation for one of the KK gluons, we end with an effective symmetry factor of $1/2$.  Eq.~\eqref{3bodyG} applies both in the regime $\mu<1/2$, when the $t\bar{t}$ pair comes from a decay of an off-shell KK gluon, and in the case $\mu>1/2$, where this process can be thought of as a two-body decay 
$g^*\to g^1g^1$, with one of the KK gluons then decaying to $t\bar{t}$. In this latter case, a narrow-width approximation for the decayed KK gluon yields
\beqa
  h_G &\approx& \sqrt{1 - 4 \mu^2} \biggl[ \biggl( \frac{g_1}{g_s} \biggr)^2 \bigl( 1 - 3 \mu^2 + 6\mu^4 \bigr) \CR
  & &+ \biggl( \frac{g_1 g_2 \LIR^2}{g_s^2 M^2} \biggr) \frac{20}{3} \bigl( 1 - \mu^2 \bigr) + \biggl( \frac{g_2 \LIR^2}{g_s M^2} \biggr)^2 \frac{1}{12\mu^4} \bigl( 1 + 12\mu^2 +56\mu^4\bigr) \biggr] \,, 
\eeqa{2bodyG}
in agreement with the direct two-body decay calculation. In the direct calculation, the symmetry factor of 1/2 mentioned above arises due to the two identical particles in the final state.

The function $h_F$ is given by
\beqa
 h_F \bigl( \tilde{g}_L, \tilde{g}_R, \mu, \gamma \bigr) &=&  \frac{1}{3\pi} \frac{\gamma}{\mu^3} \int_{2\mu}^{1+\mu^2} \ud x \frac{(x^2 - 4\mu^2)^{3/2} (1-x+\mu^2)}{(1 - x)^2 + \mu^2 \gamma^2} \CR
 & & \times
 \biggl[ \biggl(\frac{\tilde{g}_L}{g_s} \biggr)^2 \bigl( 1 - x + \mu^2 \bigr) \bigl( -2 x^2 + 5x - 2\mu^2 \bigr) \CR
 & & \qquad + 20 \biggl( \frac{\tilde{g}_L \tilde{g}_R}{g_s^2} \biggr) \mu^2 \bigl( 1 - x + \mu^2 \bigr) %\CR
%& & \hskip2cm 
+ \biggl(\frac{\tilde{g}_R}{g_s} \biggr)^2 \mu^2 \bigl( -2 x^2 + 5x - 2\mu^2 \bigr) \biggr] .\CR
\eeqa{3bodyF}
Note that the KK quark width is proportional to $m_F^2/m_{W,Z}^2$, and we have neglected subleading corrections in the vector boson mass in eq.~\leqn{3bodyF}. As before, in the case $\mu>1/2$, we may use the narrow-width approximation
\beq
  h_F  \,\approx\, \bigl( 1 - 4\mu^2 \bigr)^{3/2} \biggl[ \biggl( \frac{\tilde{g}_L^2 + \tilde{g}_R^2}{g_s^2} \biggr) \biggl( 1 -\frac{2}{3} \mu^2 \biggr) + \frac{20}{3} \biggl( \frac{\tilde{g}_L \tilde{g}_R}{g_s^2} \biggr) \mu^2 \biggr] \,, \CR
\eeq{2bodyF}
which is in agreement with an explicit evaluation of $g^* \to f^1 \bar{f}^1$. Note, however, that $f^1$ tends to be a rather broad resonance, which limits the numerical accuracy of the narrow-width approximation in this case.

\end{appendix}

\end{document}